**Asymmetries in core collapse supernovae revealed by maps of radioactive titanium in Cas A**

B. W. Grefenstette[1], F. A. Harrison[1], S. E. Boggs[2], S. P. Reynolds[3], C. L. Fryer[4], K. K. Madsen[1], D. R. Wik[5], A. Zoglauer[2], C. I. Ellinger[6], D. M. Alexander[7], H. An[8], D. Barret[9,10], F. E. Christensen[11], W. W. Craig[2,12], K. Forster[1], P. Giommi[13], C. J. Hailey[14], A. Hornstrup[11], V. M. Kaspi[8], T. Kitaguchi[15], J. E. Koglin[16], P. H. Mao[1], H. Miyasaka[1], K. Mori[14], M. Perri[13,17], M. J. Pivovaroff[12], S Puccetti[13,17], V. Rana[1], D. Stern[18], N. J. Westergaard[11], W. W. Zhang[5]

[1]Cahill Center for Astrophysics, 1216 E. California Blvd, California Institute of Technology, Pasadena, CA 91125, USA
[2]Space Sciences Laboratory, University of California, Berkeley, CA 94720, USA
[3]Physics Department, NC State University, Raleigh, NC 27695, USA
[4]CCS-2, Los Alamos National Laboratory, Los Alamos, NM 87545, USA
[5]NASA Goddard Space Flight Center, Greenbelt, MD 20771, USA
[6]Department of Physics, University of Texas at Arlington, Arlington, TX 76019, USA
[7]Department of Physics, Durham University, Durham DH1 3LE, UK
[8]Department of Physics, McGill University, Rutherford Physics Building, Montreal, Quebec H3A 2T8, Canada
[9]Université de Toulouse, UPS-OMP, IRAP, Toulouse, France
[10]CNRS, Institut de Recherche en Astrophysique et Planétologie, 9 Av. colonel Roche, BP 44346, F-31028 Toulouse Cedex 4, France
[11]DTU Space, National Space Institute, Technical University of Denmark, Elektrovej 327, DK-2800 Lyngby, Denmark
[12]Lawrence Livermore National Laboratory, Livermore, CA 94550, USA
[13]ASI Science Data Center, Via del Politecnico snc I-00133, Roma, Italy
[14]Columbia Astrophysics Laboratory, Columbia University, New York, NY 10027, USA
[15]RIKEN, Nishina Center, 2-1 Hirosawa, Wako, Saitama, 351-0198, Japan
[16]Kavli Institute for Particle Astrophysics and Cosmology, SLAC National Accelerator Laboratory, Menlo Park, CA 94025, USA
[17]INAF – Osservatorio Astronomico di Roma, via di Frascati 33, I-00040 Monteporzio, Italy
[18]Jet Propulsion Laboratory, California Institute of Technology, Pasadena, CA 91109, USA

**Asymmetry is required by most numerical simulations of stellar core collapse explosions, however the nature differs significantly among models. The spatial distribution of radioactive $^{44}$Ti, synthesized in an exploding star near the boundary between material falling back onto the collapsing core and that ejected into the surrounding medium[1], directly probes the explosion asymmetries. Cassiopeia A is a**

young[2], nearby[3], core-collapse[4] remnant from which $^{44}$Ti emission has previously been detected[5-8], but not imaged. Asymmetries in the explosion have been indirectly inferred from a high ratio of observed $^{44}$Ti emission to that estimated from $^{56}$Ni[9], from optical light echoes[10], and by jet-like features seen in the X-ray[11] and optical[12] ejecta. Here we report on the spatial maps and spectral properties of $^{44}$Ti in Cassiopeia A. We find the $^{44}$Ti to be distributed non-uniformly in the un-shocked interior of the remnant. This may explain the unexpected lack of correlation between the $^{44}$Ti and iron X-ray emission, the latter only being visible in shock heated material. The observed spatial distribution rules out symmetric explosions even with a high level of convective mixing, as well as highly asymmetric bipolar explosions resulting from a fast rotating progenitor. Instead, these observations provide strong evidence for the development of low-mode convective instabilities in core-collapse supernovae.

$^{44}$Ti is produced in Silicon burning in the innermost regions of the material ejected in core-collapse supernovae, in the same processes that produce iron and $^{56}$Ni[13]. The decay of radioactive $^{44}$Ti (in the decay chain $^{44}$Ti->$^{44}$Sc->$^{44}$Ca) results in three lines of roughly equal intensity at 67.86, 78.36 and 1157 keV. Previous detections by *COMPTEL*[5] of the 1157 keV line and *Beppo-SAX*[6], *RXTE*[7] and *INTEGRAL*[8] of the 68 and 78 keV lines were of relatively low statistical significance individually but when combined[8] they indicate a flux in each of the 67.86 and 78.36 keV lines of $(2.3 +/- 0.3) \times 10^{-5}$ ph cm$^{-2}$ s$^{-1}$. For an explosion date of 1671[2], distance of 3.4 kpc[3], and using the half life of 60 years[14], this translates into a synthesized $^{44}$Ti mass of $1.6\,^{+0.6}_{-0.3} \times 10^{-4}$ M$_{Sun}$. Due to limited spectral and

spatial resolution, previous observations are not able to constrain the line centroid or spatial distribution within the remnant, although the non-detection of the 1157 keV line by *INTEGRAL*/SPI has been used to place a lower limit of 500 km/s on the line width.

The space-based *NuSTAR* (Nuclear Spectroscopic Telescope ARray) high-energy X-ray telescope[15], which operates in the band from 3 – 79 keV, observed Cassiopeia A, the remnant of a Type IIb supernova[4], for multiple epochs between August 2012 and June 2013 with a total exposure of 1.2 Msec (Table ED1). The spectrum (Fig. 1) shows two clear, resolved emission lines with centroids redshifted by ~0.5 keV relative to the rest frame $^{44}$Ti decays of 67.86 and 78.36 keV. The telescope optics response cuts off at 78.39 keV (due to the Pt K-edge in the reflective coatings), which may affect the measured line centroid, width, and flux of the 78.36 keV line, so we focus on the 67.86 keV line for quantitative analysis. All errors are given at 90% confidence unless otherwise stated. We measure a line flux of 1.51 +/- 0.31 x $10^{-5}$ ph cm$^{-2}$ s$^{-1}$, implying a $^{44}$Ti yield of (1.25 +/- 0.3) x $10^{-4}$ M$_{Sun}$. This confirms previous spatially-integrated $^{44}$Ti yield measurements with a high statistical significance (see Supplemental Information). The $^{44}$Ti line is redshifted by 0.47 +/- 0.21 keV, corresponding to a bulk line-of-sight Doppler velocity of 1100 - 3000 km s$^{-1}$. The line is also broadened with a Gaussian half width at half maximum (HWHM) of 0.86 +/- 0.26 keV. Assuming a uniformly expanding sphere the corresponding velocity for the fastest material is 5350 +/- 1610 km/s.

The spatial distribution of emission in the 65 – 70 keV band (Fig. 2, Figure ED1) shows that the $^{44}$Ti is clumpy and is slightly extended along the 'jet' axis seen in in the X-ray

Si/Mg emission[11] and fast moving optical knots[12]. There are also knots of emission clearly evident off the 'jet' axis. There is no evident alignment of the emission opposite to the direction of motion of the compact central object (CCO) as might be expected if the CCO kick involves an instability at the accretion shock[16].

We find that at least 80% (Figure ED2) of the observed $^{44}$Ti emission is contained within the reverse shock radius as projected on the plane of the sky. Assuming a ~5000 km/s expansion velocity from above and an age of 340 years, the fastest-moving, outermost material with the highest line-of-sight velocity is 1.8 +/- 0.5 pc from the center of the explosion, which is consistent with the 1.6 pc radius estimated for the reverse shock[17]. This rules out the possibility that the $^{44}$Ti is elongated along the line of sight and exterior to the reverse shock and is only observed in the interior of the remnant due to projection effects. We conclude that a majority of the $^{44}$Ti is in the un-shocked interior.

A striking feature of the *NuSTAR* $^{44}$Ti spatial distribution is the lack of correlation with the Fe-K emission measured by *Chandra* (Fig. 3). In a supernova explosion, incomplete Si burning produces ejecta enriched with a range of elements including Si and Fe, while 'pure' iron ejecta result either from complete Si burning or from the a-rich freezeout process that also produces $^{44}$Ti. While the fraction of Fe in 'pure' ejecta is difficult to constrain observationally[18], most models predict that a significant fraction of the Fe is produced in close physical proximity to the $^{44}$Ti. Some correlation would therefore be expected. The simplest explanation for the lack of correlation is that much of the Fe-rich ejecta have not yet been penetrated by the reverse shock and therefore do not radiate in the X-ray band. While the X-rays from $^{44}$Ti decay are produced by a nuclear transition and directly trace the distribution of synthesized material, the Fe X-ray emission results from an atomic transition and traces the product of Fe density with the density of shock-heated electrons; without the hot electrons, the Fe will not be visible in the X-rays. A possible explanation of our observations is that the bulk of the Fe ejecta in Cassiopeia A have not yet been shock heated, further constraining models[18-20] of the remnant as well as the total amount of iron. An alternative explanation is that most of the Fe is already shocked and visible and some mechanism decouples the production of $^{44}$Ti and Fe and produces the observed uncorrelated spatial map.

Un-shocked or cool dense material (material that either was never heated, or has already cooled after being shock heated) might still be visible in the optical or infrared. The *Spitzer* space telescope observes line emission from interior ejecta primarily in [Si II] but it appears that there is not a significant amount of Fe present in these regions[21]. However, if un-shocked ejecta are of sufficiently low density or the wrong ionization states, then they will be invisible in the IR and optical. Low-density Fe-rich regions may in fact exist interior to the reverse shock radius as a result of inflation of the emitting material by radioactivity (the "nickel bubble" effect[22]).

The concentration of Fe-rich ejecta inferred from maps in X-ray atomic transitions is well outside the region where it is synthesized, and not in the center of the remnant interior to the reverse shock. This observation has been used to suggest the operation of a strong instability similar to that proposed for SN 1993J[23]. The presence of a significant fraction of the $^{44}$Ti interior to the reverse shock and the implied presence of interior 'invisible' iron requires this conclusion be revisited.

The measured $^{44}$Ti line widths and distribution can directly constrain mixing in the supernova engine. As evidenced by SN1987A, mixing due to Rayleigh Taylor instabilities occurring between the explosion's forward and reverse shocks (distinct from the remnant's forward and reverse shocks) may be important in some types of supernova explosions[24]. Since $^{44}$Ti is a good spatial tracer of $^{56}$Ni in all established supernova models, we can compare the measured velocity width to that predicted for $^{56}$Ni by simulations. We find that the ~5000 km/s maximum velocity and the level of Doppler line broadening compares well to Type IIb models including mixing[25] and excludes models without the growth of significant instabilities.

The evidence for asymmetries in the supernova explosion mechanism has grown steadily over the last decades. Asymmetries are implied by a number of observations[26]: the extensive mixing implied in nearby supernovae (e.g. SN 1987A); the high space velocities of neutron stars; and the polarization of supernova emission. Although different external processes could separately explain each of these observations, it is generally assumed that the asymmetries arise in the explosion mechanism. Within the convectively-enhanced supernova engine paradigm a number of mechanisms have been proposed[27]: asymmetric collapse, asymmetries caused by rotation, and asymmetries caused by low-mode convection. Of these, rotation and low-mode convection have received the most attention. Rotation tends to produce bipolar explosions along the rotation axis where the ejecta velocities are two to four times greater along this axis[28] than in the rest of the ejecta. Low-mode convection, including the standing accretion shock instability (SASI), will produce a bipolar explosion in fast-rotating stars, but is likely to produce higher order modes in slowly-rotating systems[29].

To further understand the nature of the observed $^{44}$Ti non-uniformity, we compare the observations to three-dimensional models of normal core collapse supernovae using a progenitor designed to produce the high $^{44}$Ti/$^{56}$Ni ratio needed to match the estimated yields in the Cas A remnant. We simulate two explosions that represent the extremes of explosion asymmetry: a spherically symmetric explosion and an explosion representing a fast-rotating progenitor with artificially-induced bipolar asymmetry where the explosion velocity in a thirty-degree half-angle cone near the rotation axis is increased by a factor of four relative to the rest of the ejecta. The simulated $^{44}$Ti maps (Figure ED3) indicate that the level of observed non-uniformity in Cassiopeia A is far greater than what can be produced by the spherically symmetric explosion and that the bipolar explosion (where the bulk of the fast $^{44}$Ti remains within 30 degrees of the rotation axis) cannot reproduce the observed off-axis $^{44}$Ti knots. This argues against fast-rotating progenitors as well as jet-like explosions, which are even more collimated than the bipolar explosions. The supernova is better described by an intermediate case, where the observed non-uniformity in the $^{44}$Ti is the result of a multi-modal explosion such as predicted in both low-mode Rayleigh-Taylor models[29] and Standing Accretion Shock Instability models[30]. The Cassiopeia A remnant provides the first strong evidence that this low-mode convection must occur.

**Acknowledgements** This work was supported under NASA No. NNG08FD60C, and made use of data from the Nuclear Spectroscopic Telescope Array (*NuSTAR*) mission, a project led by Caltech, managed by the Jet Propulsion Laboratory, and funded by the National Aeronautics and Space Administration. We thank the *NuSTAR* Operations, Software and Calibration teams for support with execution and analysis of these observations.

**Author Contributions**. **BG** : reduction and modeling of the *NuSTAR* Cas A observations, interpretation, manuscript preparation. **FH:** NuSTAR PI, observation planning, interpretation of results and manuscript preparation. **SB:** interpretation, manuscript review**, SR:** interpretation, manuscript preparation and review **CF:** interpretation of results, manuscript review. **KM:** observation planning, data analysis, manuscript review**, DW:** background modeling**,** data analysis, manuscript review **AZ:** background modeling, manuscript review **CE:** supernova simulations, manuscript review **HA:** image deconvolution**,** manuscript review **TK:** detector modeling, data analysis, manuscript review **HM, VR, PM:** detector production, response modeling, manuscript review**, MP:** optics calibration, manuscript review **SP,MP**: analysis software, calibration, manuscript review. **KF:** observation planning, **FC:** optics production and calibration, manuscript review, **WC:** optics and instrument production and response, observation planning, manuscript review **CH:** optics production and response, interpretation, manuscript review **JK:** optics production and response, manuscript review **NW:** manuscript review, calibration, **WZ:** optics production and response, manuscript review, **DA, DB, PG, AH, VK, DS**: science planning, manuscript review

**Author Information** Reprints and permissions information is available at www.nature.com/reprints. Correspondence and requests for materials should be addressed to bwgref@srl.caltech.edu and fiona@srl.caltech.edu

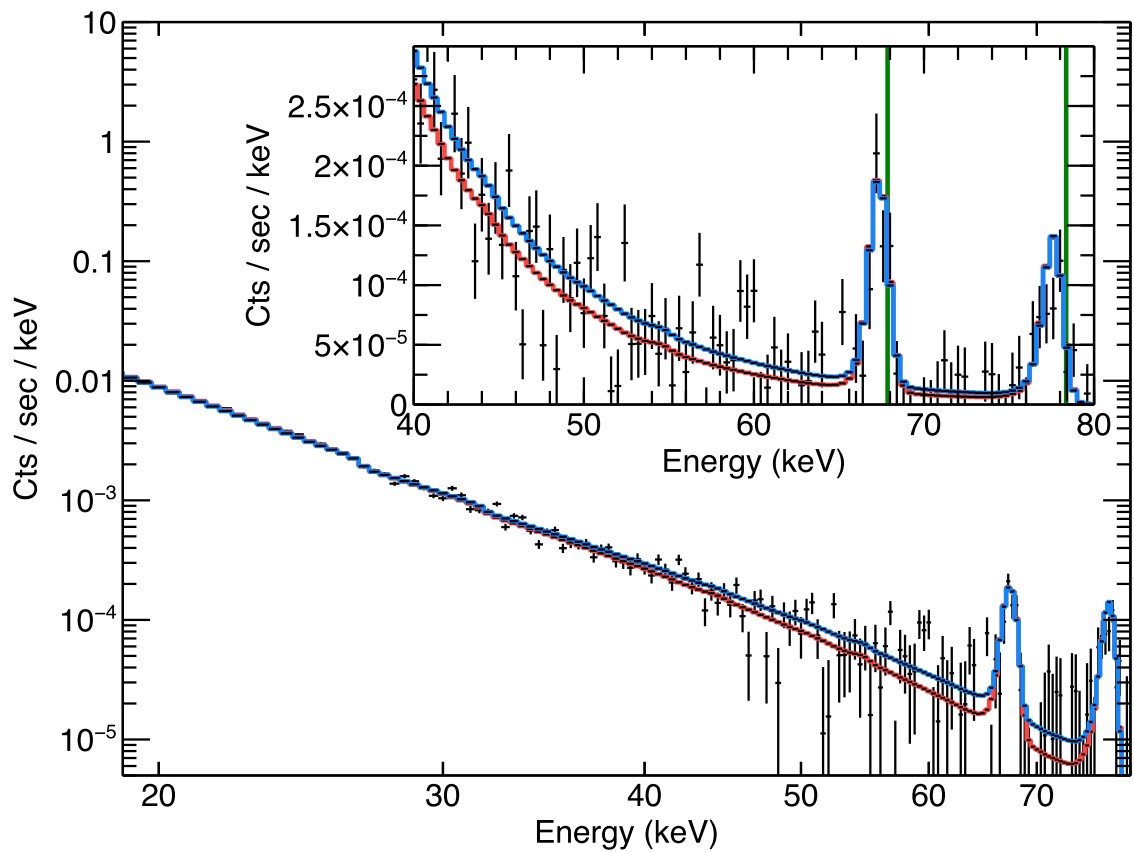

**Figure 1**: **The broadband hard X-ray spectrum of Cassiopeia A.** Data from both telescopes over all epochs are combined and shown as the grey data points with 1-$\sigma$ error bars. The spectra are shown combined and rebinned for plotting purposes only. Also shown are the best-fit continuum models for a power law (blue) and a model that describes electron cooling due to synchrotron losses (red). The continuum fits were obtained using the 10 - 60 keV data and extrapolated to 79 keV with the best-fit values for the continuum models provided in Table ED2, though the choice of continuum model does not significant affect the measurement of the lines (see Supplemental Information for details).

**Inset**: Zoomed region around the $^{44}$Ti lines showing the data and the two models on a linear scale. The vertical green lines are the rest-frame energies of the $^{44}$Ti lines (67.86 keV and 78.36 keV). A significant shift of ~0.5 keV to lower energy is evident for both lines, indicating a bulk line-of-sight velocity away from the observer. Details of the data analysis, including a discussion of the *NuSTAR* background features (Figure ED4), are given in the Supplemental Information. When the continuum is extrapolated to 79 keV there are clearly visible line features (Figure ED5) near the $^{44}$Ti line energies. Table ED3 lists the parameters of the best-fit Gaussian models of these features with the error estimates described in the Supplemental Information.

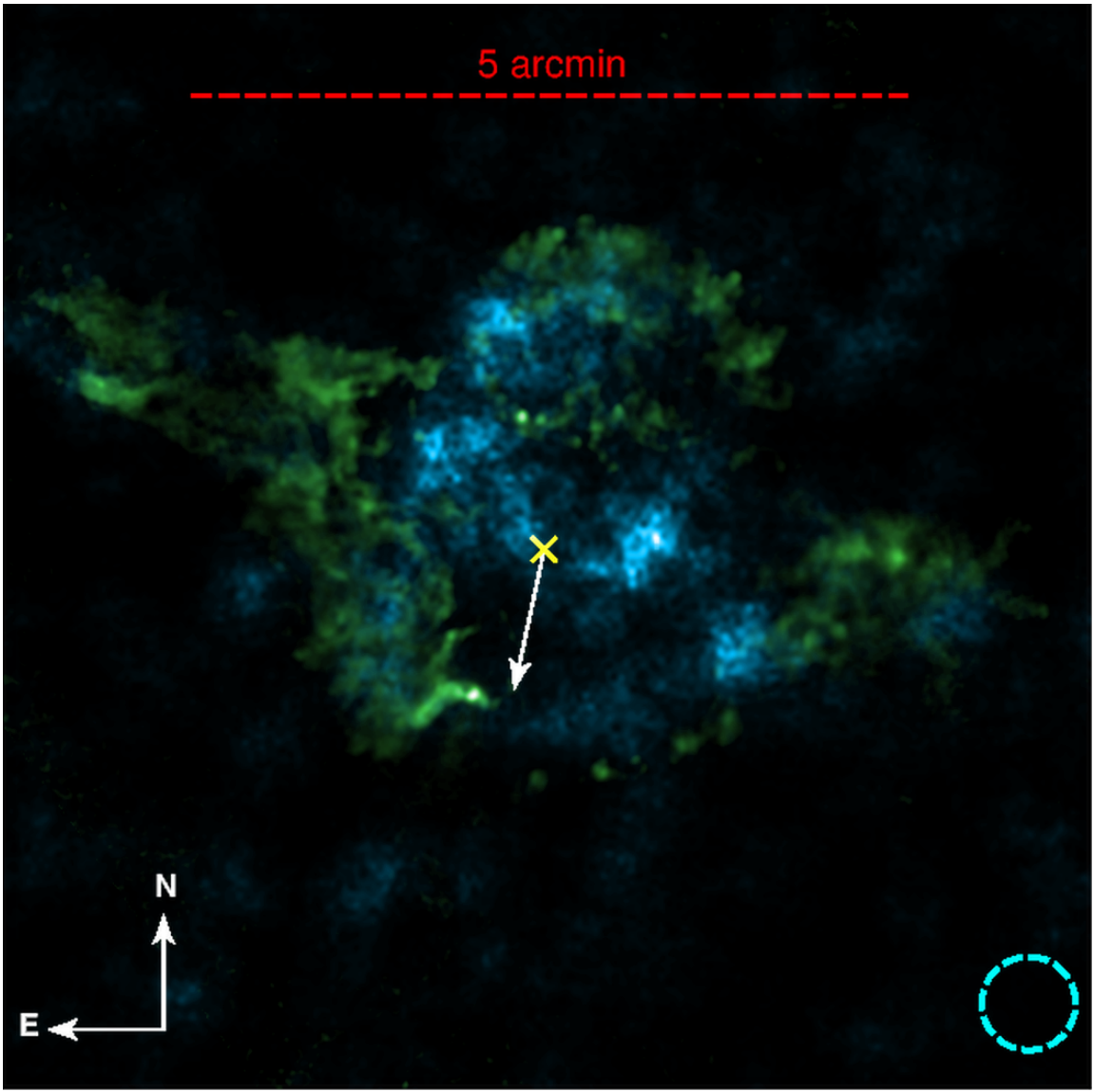

**Figure 2: A comparison of the spatial distribution of the $^{44}$Ti with the known "jet" structure in Cassiopeia A.** The image is oriented in standard astronomical coordinates as shown by the compass in the lower left and spans just over 5 arcminutes. The $^{44}$Ti observed by *NuSTAR* is shown in blue, where the data have been smoothed by a top-hat function with a radius shown in the lower right (dashed circle). The $^{44}$Ti is clearly resolved into distinct knots and is non-uniformly distributed and almost entirely contained within the central 100 arcseconds (see the Supplemental Information and

Figure ED2). Shown for context in green is the *Chandra* ratio image of the Si/Mg band (data courtesy NASA/CXC, Si/Mg ratio image courtesy J. Vink), which highlights the jet/counterjet structure, the center of the expansion of the explosion[2] (yellow cross), and the direction of motion of the compact object (white arrow). In contrast to the bipolar feature seen in the spatial distribution of Si ejecta that argues for fast rotation or a jet-like explosion, the distribution of $^{44}$Ti is much less elongated and contains knots of emission off of the jet axis. A reason for this may be that the Si originates from the outer stellar layers and is likely highly influenced by asymmetries in the circumstellar medium, unlike the $^{44}$Ti, which is produced in the innermost layers near the collapsing core.

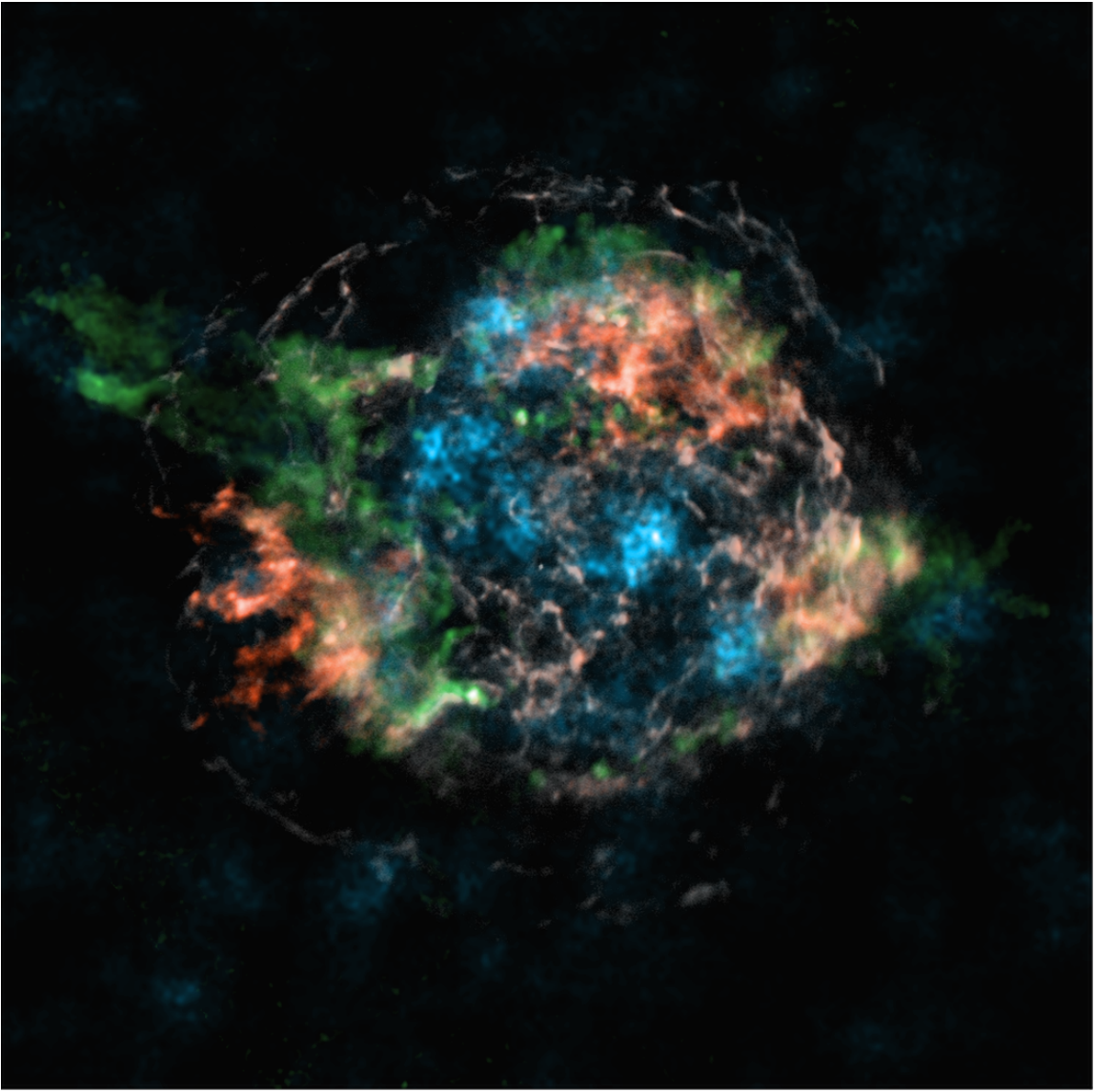

**Figure 3: A comparison of the spatial distribution of $^{44}$Ti with known Fe K emission in Cassiopeia A.** We reproduce the spatial distributions shown in Fig 2 and add the 4 to 6 keV continuum emission (white) and the spatial distribution of X-ray bright Fe (red) seen by *Chandra* (Fe distribution courtesy U. Hwang). We find that the $^{44}$Ti does not follow the distribution of Fe-K X-ray emission, suggesting either that a significant amount of Fe remains un-shocked and therefore does not radiate in the X-ray, or the Fe/Ti ratio in the ejecta deviates from the expectation of standard nucleosynthesis models.

**Extended Methods**

**Section 1: Observations and preliminary data reduction:**

Cas A was observed by *NuSTAR* during August and November of 2012 and March and June of 2013 for a total of over 1.2 Ms of exposure (Table ED1). We reduced the *NuSTAR* data with the *NuSTAR* Data Analysis Software (NuSTARDAS) version 1.2.0 and *NuSTAR* CALDB version 20130509. The NuSTARDAS pipeline software and associated CALDB files are fully HEASARC ftool compatible and are written and maintained jointly by the ASI Science Data Center (ASDC, Italy) and the *NuSTAR* Science Operations Center (SOC) at Caltech. The NuSTARDAS pipeline generates Good Time Intervals (GTIs) for each observation that exclude periods when the source is occulted by the Earth and when the satellite is transiting the South Atlantic Anomaly (SAA), a region of high particle background. The default "depth cut" is applied to reduce the internal background at high energies.

**Section 2: Background Estimation and Modeling:**

Cassiopeia A is a bright, extended source so that that there are no regions in the field-of-view of the telescope that can be used to directly estimate the background in the source region. Instead we model the background and produce background images and spectra.

The *NuSTAR* background is well described by a non-focused Cosmic X-ray Background (CXB) component that dominates at low energies (<20 keV), a focused CXB component, and an internal background, composed of a continuum along with many lines due to radioactive decay in the instrument and fluorescence from the CsI shield, that dominates at higher energies (>20 keV)[31]. Long, *NuSTAR* deep field observations from the Extended *Chandra* Deep Field South (ECDFS) and Cosmological Evolution Survey (COSMOS) observations were used to determine the spatial variation of the CXB across each focal plane and the relative strength of the internal background for each detector. Typically, background regions free of source counts are used to determine the normalization of the CXB components and the internal background component (including the relative strengths of the internal background lines that may show some long term variation related to activation in the instrument and spacecraft). We use the normalizations of the various components to produce background images and to predict the background spectrum anywhere on the detector (e.g. in the source region). In the case of Cas A, we add an additional phenomenological model to account for the redistribution of source counts into the background regions via the PSF wings (Figure ED4).

Background spectra are generated for the source region for each observation epoch (excluding the contribution from the phenomenological model) using the fakeit command in XSPEC with the exposure set to be the same as for the source exposure. This results in a Poisson-distributed realization of the background spectrum in the source region for each observation. We address the implications of this below.

Background images are produced for the 65-to-70 keV band using the normalization of the internal background (the CXB is negligible in this band) and the known detector-to-detector amplitude variation. The background images produced by our model account for spatial variations in diffuse components as well as focal plane efficiency variations. We combine the background images, weighting them by the effective exposure.

There is both Pt and W in the multilayer coatings on the optics, which could give rise to a source-dependent fluorescence lines near the $^{44}$Ti lines for sources with spectra that extend above the Pt K-edge. However, we do not see fluorescence lines appearing in the spectra of targets used for the calibration of the optics, such as bright X-ray binaries or the Crab Nebula. We infer that any fluorescence from the multilayer coatings is too diffuse to observe.

**Section 3: Spatial Distribution of the Ti-44 in Cas A:**

To generate the significance maps we convolve the mosaicked counts image and the accompanying background image with a top-hat function of radius 20 arcseconds (8 image pixels). We calculate the probability of detecting the observed number of counts or higher from a Poisson fluctuation in the background. A region containing source counts would then have a low Poisson probability of being a random fluctuation in the background[32]. For the $^{44}$Ti Cas A image we generate contours at probability levels of 2.7e-3, and 6.3e-5, corresponding to background fluctuations of 3 and 4 sigma (Figure ED1). In the 65 to 70 keV band we find 5721 counts interior to 150 arcseconds, while our background model predicts 4780 counts. This constitutes a 13.6-sigma deviation from the expected background, constituting a highly significant detection. However, this detection cannot distinguish between any continuum or line contribution over this band. The significance of the line is discussed below.

To estimate the radial distribution of the $^{44}$Ti we compute the number of counts contained in annuli with successively larger radii measured from the optical center of the remnant (RA, DEC) of ($23^h23^m27^s.77, +58°48'49".4$, J2000)[33], hereafter referred to as the "center of the remnant". We do the same operation with the background images to determine the expected background. After subtracting the background we find that more than 80% of the enclosed flux lies within 100 arcseconds of the center of the remnant (Figure ED2). The effects of the optics point spread function (PSF) and the vignetting (the loss of effective area as a function of off-axis angle in the optics) on the radial distribution are not included here. Both of these effects depend on the (unknown) intrinsic source distribution, but neither should be a strong enough effect to alter the statement that a majority of the photons are found inside of 100 arcseconds.

**Section 4: Spectroscopy**

We extracted a 120-arcsecond radius region centered on the optical center of Cassiopeia A. We extracted source spectra, ARFs, and RMFs for each observation using the nuproducts FTOOL with the "extended source" option. All 10 spectra (five epochs and two focal plane modules, or FPMs) were simultaneously fit in XSPEC. Because of the low number of source counts, we use the "W-stat" as the fit statistic in XSPEC and the Pearson's chi-square as the test statistic. The background subtraction method is described below.

**Section 4-1: Determination of the Underlying Continuum**

Previous hard X-ray observations of Cas A have shown that the hard X-ray continuum can be described either by the "powerlaw" or "srcut" standard spectral models in XSPEC. We make a slight modification to the standard srcut model to account for the softer radio photon index found in Cassiopeia A (0.77[34] compared to 0.5, which is typical for srcut).

While a detailed analysis of the underlying hard X-ray continuum is beyond the scope of this paper and will be addressed in future work, here we must determine how the choice of continuum model affects the estimates of the $^{44}$Ti lines. To do this, we first fit both models to the spectra over a 15 to 60 keV bandpass. In this range the Poisson fluctuations of the background are negligible, so we use a single realization of the background spectrum. To compare the continuum models we use the "goodness" command in XSPEC to generate 1,000 realizations of the model parameters based on the covariance matrix and compute the test statistic for each of these model realizations. On average, 50% of the model realizations should produce test statistic values greater than the best-fit test statistic value. If many (significantly more than 50%) of the simulated spectra produce fits with test statistic values less than our best-fit model, then we may reject the model. Since we are not attempting a physically motivated model, but rather a good description of the continuum, we do not constrain the srcut normalization (which is given in flux at 1 GHz, typically constrained from radio observations). However, for completeness we also include a constrained srcut model with the normalization at 1 GHz fixed to be 2720 Jy to match the value used by previous authors[35]. This value is the 1 GHz flux at epoch 1977[34], though the decay rate of Cas A's 1 GHz flux of 0.6 to 0.7 percent per year[36] gives a 1 GHz flux in 2013 of about 2300 Jy.

The best-fit continuum parameters are given in Table ED2. The powerlaw model is a good fit to the continuum data and produces a better fit than 50% of the realizations, while the best-fit srcut model probides a better fit than only 10% of the realizations. This suggests that we should reject the srcut model. The constrained srcut model is comparably rejected, so we do not include either of these continuum models in our analysis below. A detailed analysis of the continuum models and their spatial dependence will be undertaken in a subsequent paper.

**Section 4-2: Models of the line parameters**

We extend the fit range to 79 keV and observe significant spectral features where we expect the $^{44}$Ti lines (Figure ED5). We add two emission lines to the continuum model using the "gauss" XSPEC model. This roughly corresponds to a spatially uniform, isotropically expanding sphere of ejecta, which we acknowledge may not be the best representation of the actual distribution. Detailed spatially resolved spectroscopy will be performed in future work using additional future observations of Cassiopeia A.

The 78.36 keV $^{44}$Ti X-ray line is near the Platinum absorption edge at 78.395[31] keV in the optics, making the observed line intensity extremely sensitive to any Doppler shift in the line centroid. To account for possible bias in the fits, we use three test cases: (1) two

Gaussian model components with the line widths, redshifts, and normalizations tied together; (2) two independent Gaussian model components to data extending to 79 keV; and (3) a single Gaussian line to the 68 keV line, ignoring the counts above 72 keV. We discuss these three cases below.

**Section 4-3: Statistical Errors**

As stated above, our background models are limited by Poisson statistics. If the background were featureless (i.e. a smooth continuum) over this region this would not affect our line fits. However, because of the background line structure near 65, 67, and 75 keV (Figure ED4, inset), Poisson fluctuations in the background can affect the best-fit parameters of the line shape, centroid, and flux. We marginalize over this effect by generating 1,000 realizations of the background for each FPM and for each epoch and finding the best-fit continuum and line parameters.

We estimate our measurement and the uncertainty in the measurement as follows. The distributions of fits over the 1000 background realizations describe the variation of best-fit parameters due to Poisson background fluctuations, so we take their mean value of each distribution to be the best estimate of the parameter. For each background realization, our ability to constrain each parameter is derived by traversing chi-square (or C-stat in this case) space by varying the fit parameters according to the co-variance matrix, which is driven by the uncertainties in the Poisson-distributed source and background spectra. We adopt the mean 90% "fit" errors as the uncertainties on our measurement.

**Section 4-4: Systematic Errors**

Case 1 produces lines that are systematically shifted to lower energy (higher redshift) and higher flux. This points to a mismatch between the estimated effect of the high-energy cutoff estimated in the optics response files and the actual effect of the Pt absorption. Any assumed relation between the lines (i.e. fixing their widths, relative centroids, or relative fluxes as we have done in Case 1) will be incorrect when compared to the data. As both Case 2 and Case 3 produce nearly identical results for the 68 keV line parameters, we choose Case 2 as our model of choice in the main text and report only the fit values for the 68 keV line below. Results for all three models are given in Table ED3 with the 90% statistical errors given. For Case 2, this represents an 8-σ detection.

The *NuSTAR* effective area has a residual uncertainty of ~10% over the 60-80 keV band-pass, which we add in quadrature with our (larger) statistical errors in the summary below.

If $^{44}$Ti is ionized to He-like states or beyond, the electron-capture rate is reduced and the mean lifetime can increase. However, our image analysis above indicates that most of the $^{44}$Ti is within the reverse shock and at low ionization. Even if most of it were already shocked, a detailed calculation[37] has shown that the current ionization state is very unlikely to be as high as required for a significant impact on the lifetime.

**Section 4-5: Summary of Spectroscopic Results**

After combining the statistical and systematic errors, we find a line flux (90% error range) at 68 keV of $1.51 +/- 0.31 \times 10^{-5}$ ph / s / cm$^2$ which corresponds to an initial $^{44}$Ti synthesized mass of $1.25 +/- 0.3 \times 10^{-4}$ M$_{Sun}$. The line is shifted with respect to its rest-frame energy of 67.86 keV and is observed at 67.39 +/- 0.21 keV, corresponding to a bulk line-of-sight Doppler velocity of 2050 +/- 950 km / sec. The line is also broadened with a Gaussian half-width half-max (HWHM) of 0.86 +/- 0.26 keV. Assuming a uniform sphere, then HWHM/0.71 describes the fastest moving material, so the width of the line corresponds to a maximum line-of-sight velocities of to 5350 +/- 1650 km / sec. We caution though that this interpretation is sensitive to the assumptions about the distribution of the ejecta, however this does place reasonable limits on how fast the ejecta could be moving along the line-of-sight.

If we assume a distance of 3.4 kpc and an age of 340 years, the furthest the ejecta could be along the line of sight is 1.8 +/- 0.5 pc. If the reverse shock is described by a spherical shell then its 100-arcsecond radius translates into a radial extent of 1.6 pc, so the bulk of the $^{44}$Ti is within the reverse shock.

**Section 5: Constraints on the Supernova Engine**

One constraint on the explosion engine is the total $^{44}$Ti yield. $^{44}$Ti is produced in Silicon burning in the innermost ejecta of core-collapse supernovae[38] and the total yield, as well as the $^{44}$Ti /$^{56}$Ni ratio is sensitive to the stellar profile and explosion energy. The $^{44}$Ti/$^{56}$Ni ratio can be tuned by different progenitors and explosion strengths, and although the high value of this ratio suggests asymmetric explosions, it is not a direct probe of the asymmetry.

The spatial distribution of the $^{44}$Ti is a much stronger tracer of the explosion asymmetry. In general, the $^{44}$Ti spatial distribution traces that of the $^{56}$Ni. However, there are a few basic aspects of $^{44}$Ti production that argue that, if anything $^{44}$Ti should be produced further out and at faster velocities than $^{56}$Ni. First, there are high-density conditions (that might occur in the innermost ejecta) where $^{56}$Ni is produced and $^{44}$Ti is produced at a low rate[34], arguing that the $^{44}$Ti/$^{56}$Ni should be lowest in the innermost ejecta. In addition, at low-densities, $^{44}$Ti can be produced in partial silicon burning where the $^{56}$Ni yield is low, producing a high $^{44}$Ti/$^{56}$Ni ratio. However, these effects tend to be small in most supernovae and so we assume in the main text that the $^{44}$Ti is a good tracer of the $^{56}$Ni yield and that the maps of the $^{44}$Ti distribution can directly constrain the asymmetry in the supernova engine.

To compare our $^{44}$Ti distributions to explosion models, we ran two simulations of Cassiopeia A (using a 23 solar mass binary progenitor with an explosion energy of $2 \times 10^{51}$ ergs[39]) using the SNSPH code[40,41]: a symmetric explosion and a bimodal explosion with a factor of 4 increase along the jet axis to match the expectations of normal rotating models. Figure ED3 shows the contour profiles of the $^{44}$Ti yield for these two explosions. Although the symmetric simulation includes extensive Rayleigh-Taylor mixing, it is unable to reproduce the observed non-uniformity in the observed $^{44}$Ti distribution. Similarly, the off-axis $^{44}$Ti argues against the extreme asymmetries produced by both fast-rotating progenitors as well as "jet" supernova engines for Cassiopeia A and argue for the presence of higher-order modes in the explosion. With the rise of the "collapsar" engine for gamma-ray bursts[42], alternate core-collapse supernova engines invoking magnetic-field produced jets have also been studied and suggested to explain remnants such as Cassiopeia A[43,44]. These jet-driven explosions produce extreme bimodal asymmetries and are similarly disfavored by our results. As neither the spherically symmetric nor the strongly asymmetric explosion reproduce all of the observed properties of the $^{44}$Ti spatial distribution, the progenitor of the Cassiopeia A remnant was most likely between these two extremes: a mildly asymmetric explosion developing low-mode convection.

44    Wheeler, J. C. & Akiyama, S. Asymmetric supernovae and gamma-ray bursts. *New Astronomy Reviews* **54**, 183-190, (2010).

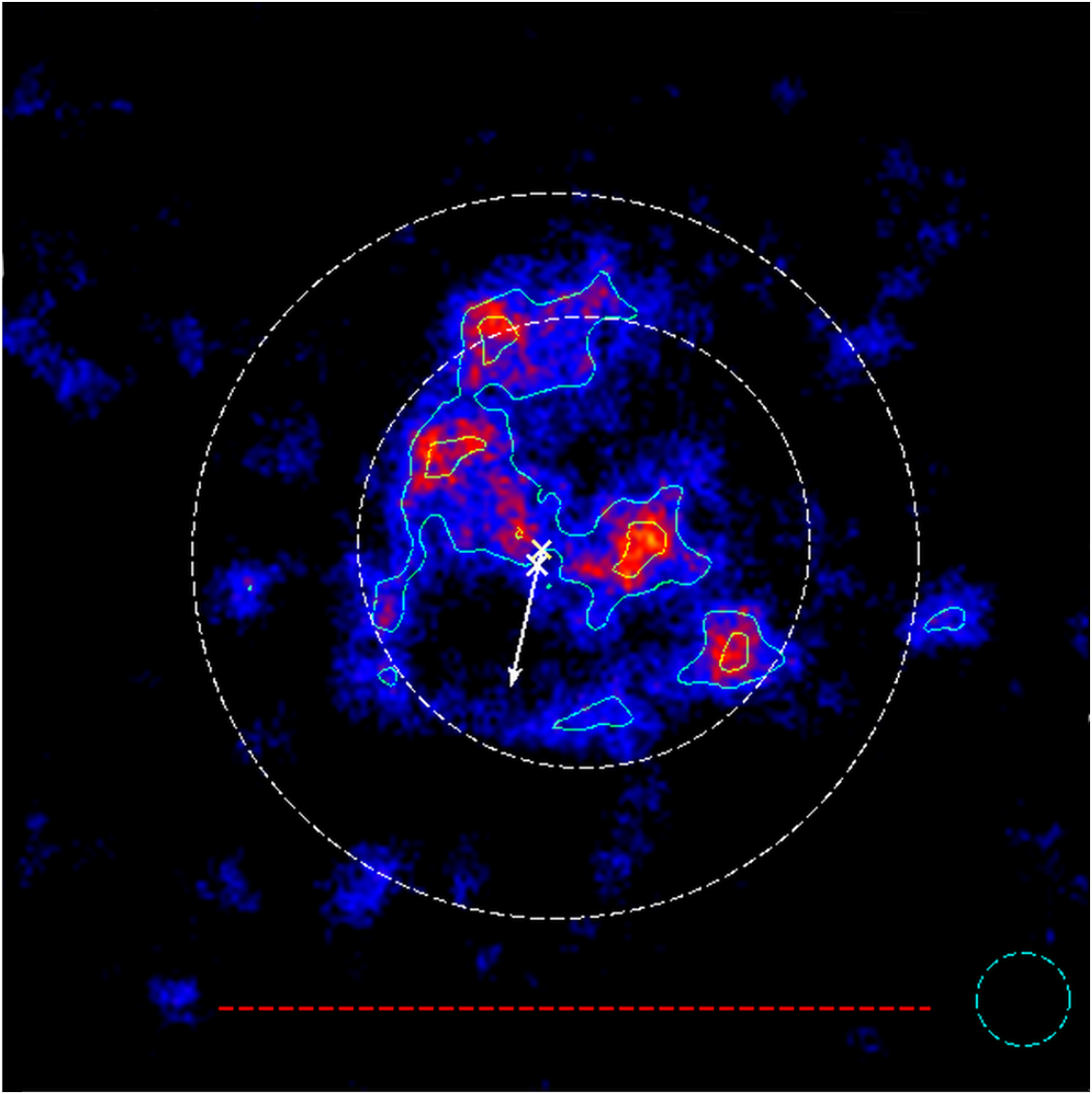

**Figure ED1: The background-subtracted image of Cas A in the 65 to 70 keV band containing the 68 keV $^{44}$Ti line showing the significance of the $^{44}$Ti knots.** The data

have been smoothed with a 20-arcsecond-radius top hat function (dashed circle) and are shown with 3 and 4 sigma significance contours (green). In addition to the features shown in Figure 1 we also show locations of the forward (R ≈ 150 arcseconds) and reverse (R ≈ 100 arcseconds) shocks[17] (white dashed circles) for context. The $^{44}$Ti clearly resolves into several significantly identified clumps that are non-uniformly distributed around the center of expansion.

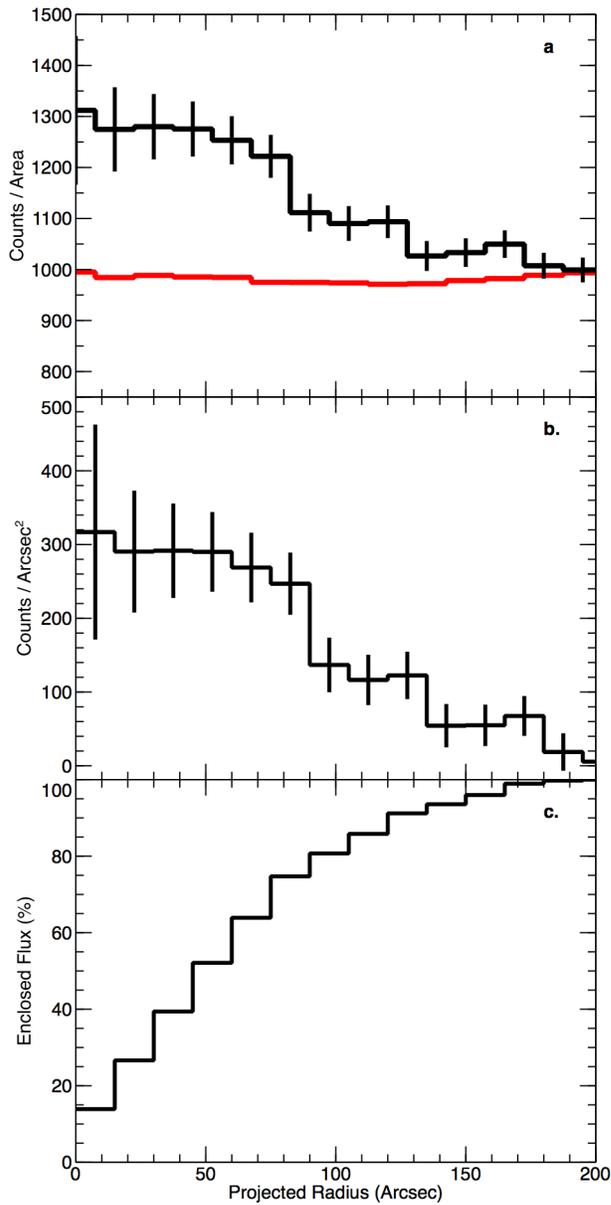

**Figure ED2: The radial profile of the $^{44}$Ti emission.** We collect each photon in annular bins of increasing radius in the plane of the sky without any spatial smoothing. **Panel A:** The radial profiles of the $^{44}$Ti the data in the 65 to 70 keV (black) and the radial profile expected from the background images (red), scaled by the area of each annulus and shown in units of counts / arcsec$^2$. **Panel B:** The background-subtracted radial profile. **Panel C:** The percent of enclosed flux in annuli of increasing radii as observed on the plane of the sky. All error bars are 1-sigma.

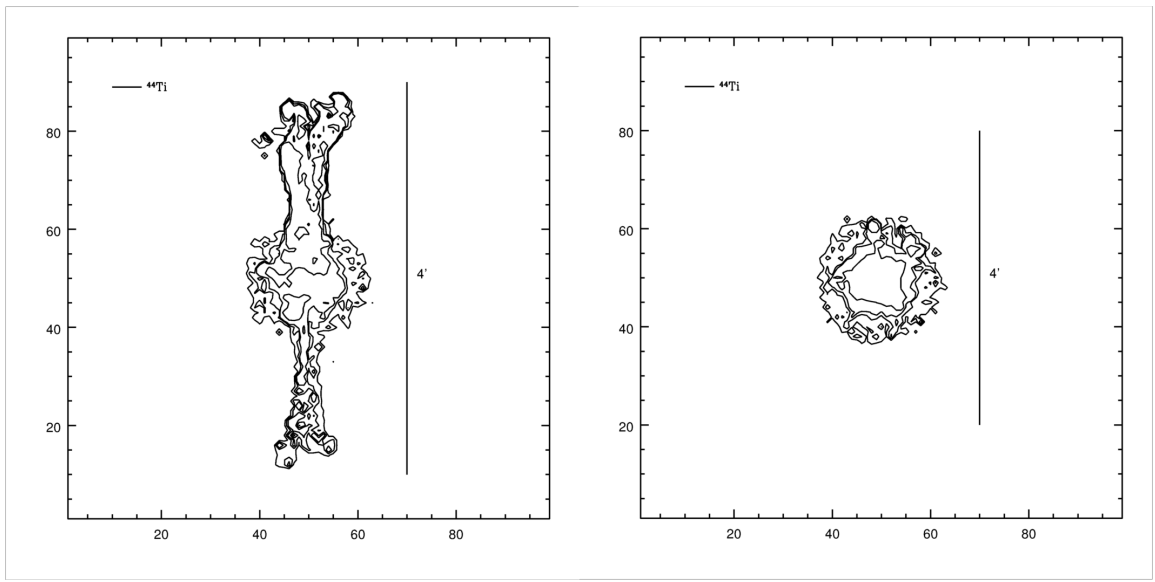

**Figure ED3**: **Simulated $^{44}$Ti intensity contours for a symmetric and a bipolar explosion.** The vertical line shows a 4' scale (note the different spatial scale between the symmetric (left) and bipolar (right) explosions). The non-uniformities in the observed $^{44}$Ti spatial distribution rule out the purely symmetric explosion, even with extensive mixing. Similarly, the presence of $^{44}$Ti outside of the "jet" axis argues against the rapidly rotating progenitor that produced the bipolar explosion. We therefore argue that the explosion that produced Cassiopeia A is somewhere between these two extremes and that this is the first clear example of a low-mode convection explosion.

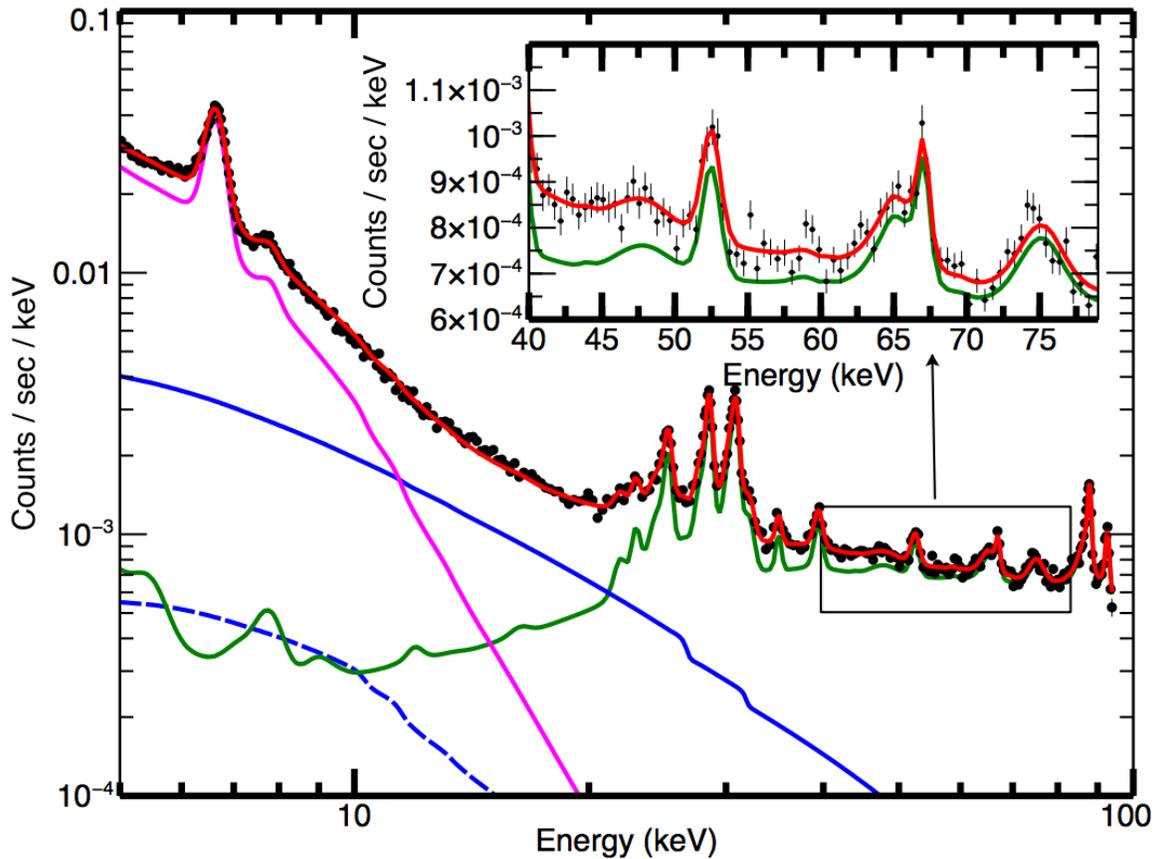

**Figure ED4: The background spectral model fit for one of the Cas A epochs.** Shown are the data from the background regions (black points with 1-sigma error bars included but not visible), the instrumental background (green), the CXB components (blue, dashed is the focused CXB component), the phenomenological "source" model (magenta), and the total background model (red). **Inset:** The background spectrum near the $^{44}$Ti emission lines showing the features that we model. The broad lines at 65 and 75 keV are likely neutron capture emission features, while the narrow line near 67 keV is an internal activation line in the CdZnTe detectors. See supplemental information for more details.

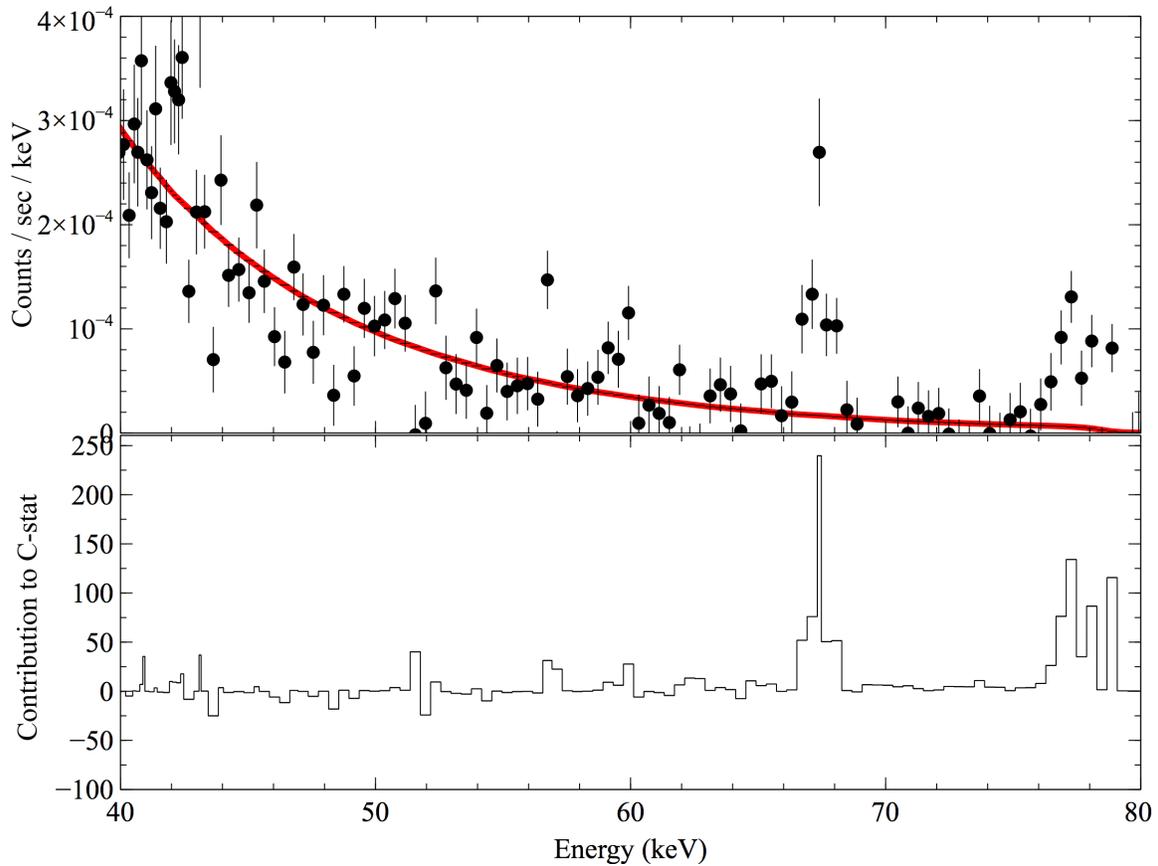

**Figure ED5: The significant signals observed in the spectrum near 68 and 78 keV.**

**Panel a:** The black points (1-sigma error bars) are the data shown after the background model spectrum has been subtracted from the source data. The red continuum is the best-fit powerlaw continuum over the 20-80 keV band-pass. **Panel b:** The contribution to the C-stat statistics for each spectral bin. The large signals near 68 and 78 keV (the $^{44}$Ti emission lines) suggest that an additional spectral component is required. See supplemental information for details.

TABLE 1

| Sequence ID | Date Start | Date End | Exposure (ks) |
|---|---|---|---|
| 40001019002 | 2012-08-18 | 2012-08-26 | 293 |
| 40021002002 | 2012-11-23 | 2012-11-29 | 288 |
| 40021002006 | 2013-03-02 | 2013-03-04 | 159 |
| 40021002008 | 2013-03-05 | 2013-03-09 | 228 |
| 40021003003 | 2013-05-28 | 2013-06-03 | 233 |
| | | Combined Exposure | 1.2 Ms |

**Table ED1: List of observations used in this analysis.** Start/stop dates are given, as has the effective exposure of each observation. The exposure has been corrected for period when the source was occulted by the earth, periods where the instrument was not taking data, and for the rate-dependent livetime of the instrument.

| Model | Index/Log(Break) | Normalization |
|---|---|---|
| Powerlaw | 3.31 (3.37 - 3.41) | 1.58 (1.50 - 1.66)* |
| SRCUT ($\alpha = 0.77$) | 17.356 (17.341 - 17.368) Hz | 1330 (1280 - 1400)** |
| SRCUT ($\alpha = 0.77$) | 17.356 (17.341 - 17.368) Hz | 2720 (fixed)** |

* Flux given in units of $10^{-5} \frac{ph}{cm^2 sec}$ at 1 keV
** Jy at 1 GHz

**Table ED2: Best-fit continuum parameters.** Results for the considered continuum models. Error ranges for the parameters are given at the 90% level.

| Model | Flux* | $\sigma$ (keV) | Measured Line Energy (keV) |
|---|---|---|---|
| Case 1 | 1.59 ± 0.28 | 0.72 ± 0.19 | 67.35 ± 0.17 |
| Case 2 | 1.51 ± 0.31 | 0.73 ± 0.22 | 67.39 ± 0.21 |
| Case 3 | 1.51 ± 0.31 | 0.73 ± 0.22 | 67.40 ± 0.21 |

* Flux given in units of $10^{-5} \frac{ph}{cm^2 sec}$

**Table ED3: Results from spectral analysis.** Results for all three spectral models for the $^{44}$Ti emission are given with 90% error estimates from the background Monte Carlo analysis. Errors are the statistical estimates from the fit parameters and do not include the systematic uncertainties in the *NuSTAR* effective area.